\documentclass[11pt]{article}

\usepackage[a4paper,top=3cm,bottom=4cm,left=3cm,right=3cm]{geometry}

\usepackage{graphicx}
\usepackage{amsmath}
\usepackage[english]{babel}

\usepackage[hidelinks=true,linkcolor=black,citecolor=black]{hyperref}
\usepackage[labelfont=bf]{caption}
\usepackage[style=ieee, doi=true]{biblatex}
\bibliography{references}
\AtBeginBibliography{\small}
\usepackage{doi}
\usepackage{csquotes}

\title{\textbf{Bridging the gap: Using deep learning to reconstruct noise-reduced super-resolved OCT images from gapped spectra}}

\author{Jonas Nienhaus\textsuperscript{1}, Thomas Schlegl\textsuperscript{1}, Wolfgang Drexler\textsuperscript{1}, Tilman Schmoll\textsuperscript{2}, \\and Rainer A. Leitgeb\textsuperscript{1,*}} 
\date{\footnotesize
    \textsuperscript{1}Center for Medical Physics and Biomedical Engineering, Medical University of Vienna, Vienna, Austria\\
    \textsuperscript{2}Carl Zeiss Meditec AG, Oberkochen, Germany\\
	\textsuperscript{*}rainer.leitgeb@meduniwien.ac.at
}

\renewenvironment{abstract}
{\noindent\large\textbf{Abstract:\space}\normalsize}
{\vspace{7pt}}

\begin{document}

\maketitle

\begin{abstract}\noindent
Fourier-domain (FD) optical coherence tomography (OCT) depends on broadband sources to maximize axial resolution and image quality.
However, these lasers significantly drive device cost or may be unavailable at desired wavelength and bandwidth ranges.
A potential solution lies in integrating multiple, more affordable sources with lower individual bandwidth into a single system.
However, difficulties arise if the resulting spectrum exhibits discontinuities.
In this letter, we present a method that can combine OCT images from a flexible number of spectra with arbitrary, possibly non-overlapping gaps using a neural network.
Compared to low-resolution input images, reconstructed B-scans are super-resolved, preserve even fine and low-contrast details and edges, and exhibit strong noise reduction that increases with the band gap.
The proposed method could thereby provide a major step towards high-quality OCT imaging using spectrally disjoint, low-bandwidth sources.
In broadband settings, it can be directly applied as a Fourier-domain masked autoencoder for self-supervised image quality enhancement.
\end{abstract}
\vspace{1em}

Fourier-domain optical coherence tomography (FD-OCT) offers depth-resolved imaging.
However, high axial resolution requires broadband sources \cite{Izatt2015TheoryOpticalCoherence}.
Such sources are a major cost contributor and market barrier, which is why recent works have investigated the use of low-cost alternatives such as thermally tuned VCSEL sources \cite{Kendrisic2023ThermallytunedVCSEL}.
While being a promising direction towards making OCT more accessible, the bandwidth of such sources remains very limited.
A potential solution may lie in the combination of multiple sources with distinct wavelength ranges.
In spectral-domain (SD) OCT, combining multiple sources such as superluminescent diodes already allows imaging at very high bandwidths \cite{Schmitt1997opticalcoherencemicroscope}.
For swept-source (SS-) OCT, phase-aligned combination is more challenging, and straight-forward processing yields only degraded images due to the non-Gaussian spectral shape.
Existing approaches either require band overlap \cite{Thakur2026Computationalframeworkcombining} or employ model-based algorithms that are computationally expensive \cite{Wang2019OpticalCoherenceTomography} or not designed for scenarios with low signal-to-noise ratio or dense sample structure \cite{Pan2026OpticalCoherenceTomography}.
In another approach working in the image domain, Lichtenegger, Salas \emph{et al.} \cite{Lichtenegger2021Reconstructionvisiblelight} used generative adversarial networks to restore high quality OCT B-scans from the distorted images.
However, while this approach works on images that are already degraded, it would be desirable to prevent these distortions in the first place.

In this letter, we propose a Fourier-domain masked autoencoder (FD-MAE) as an alternative approach to deep-learning-based combination of multiple sources which makes no restrictions on the band gap. 
Conceptually reminiscent of masked autoencoders \cite{He2022MaskedAutoencodersAre}, our model is trained via input masking, here in the spectral domain, and subsequent image reconstruction.
With this self-supervised learning, we describe for the first time a denoising effect that increases with the band gap.

Individual reconstruction of the spectra yields a set of B-scans with poor individual axial resolution (compared to an ideal combined bandwidth)
\begin{equation}
	\delta z_i = \frac{2\ln2}{\pi}\frac{\lambda_i^2}{\Delta\lambda_i} \quad i\in\lbrace1,2\rbrace,
\end{equation}
in which \(\lambda_i\) and \(\Delta\lambda_i\) are central wavelength and bandwidth (FWHM) of the \(i\)-th source \cite{Izatt2015TheoryOpticalCoherence}.
However, these scans are otherwise free of the distortions that would arise when reconstructing the combined spectrum, and together offer complementary information about sample structure.
Furthermore, assuming \(\Delta\lambda_1=\Delta\lambda_2\), the speckle correlation 
\begin{equation}\label{eq:speckle_correlation}
	\rho_{12} = \exp\left( \frac{- \left(\lambda_2 - \lambda_1\right)^2}{2\Delta\lambda_{1,2}^2} \right)
\end{equation}
of two independent sources decreases rapidly with the spectral offset \cite{Pircher2003Specklereductionoptical}.
The desired structural sample information, however, is contained along the spectrum at reduced axial resolution, if the individual signal-to-noise ratio is sufficient.


\begin{figure}[htbp]
\centering\includegraphics[width=\linewidth]{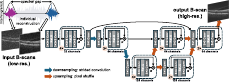}
\caption{Proposed FD-MAE architecture which combines B-scans from spectral sub-bands into a single, super-resolved image. Kernel sizes and number of feature maps follow \cite{Nienhaus2026Datacentricphysics}.}
\label{fig:network}
\end{figure}

An overview of the proposed Fourier-domain masked autoencoder (FD-MAE) is shown in Figure \ref{fig:network}.
It consists of a pre-processing step, in which the individual spectral bands are reconstructed to low-resolution input B-scans, and the combination step performed by a neural network.
The backbone network architecture is oriented on recent work on saturation artifact removal \cite{Nienhaus2026Datacentricphysics} and receives a single set of low-resolution input B-scans and outputs a single super-resolved tomogram.
Input images are first processed by 3D convolutions as pseudo-slabs, for which they need to have equal pixel dimensions. 
In our experiments, this was ensured directly by selecting equally-sized spectral windows, but it may also be achieved e.g. via resampling or zero-padding during reconstruction.
Similarly, we chose a final axial upscaling factor of 4 in accordance with the selected window widths. This factor could be modified by only adjusting the final upsampling operation.


\begin{figure}[htbp]
\centering\includegraphics[width=0.9\linewidth]{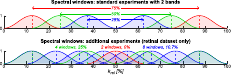}
\caption{Spectral windows and gaps (distance of centers in \(k_\mathrm{rel} [\%]\)) used for input B-scans in different experiments.}
\label{fig:windows_plot}
\end{figure}

We evaluated the proposed method on three datasets with available interferograms, in which various band gaps were introduced artificially.
For all three data sets, interferograms were windowed in \(k\)-space via Hann-windowing with widths that correspond to 25\% of the entire \(k\)-range to obtain gapped spectra, as shown in Figure \ref{fig:windows_plot}.
In particular, for all datasets we evaluated FD-MAE with three band gaps, which we denote by the distance of the central wavenumbers relative to the full bandwidth in percent: adjacent non-overlapping windows (25\%), an intermediate gap (50\%), and the largest possible gap (75\%).   
B-scans reconstructed from the full spectra served as ground truth.
As is common in super-resolution tasks, models were trained end-to-end to minimize mean absolute error between output and this target.
Batch sizes and number of epochs were chosen individually depending on dataset and image size.
Models were trained on an Nvidia RTX 3090 GPU with 24 GB of memory using an initial learning rate of \(10^{-4}\) with cosine annealing scheduling.

On the porcine anterior segment OCT (PASO) dataset \cite{Nienhaus2025PASOMultipurposePorcineARTICLE}, we trained the model on 56832 training B-scans from 111 volumes, while reserving 30 independent volumes with 15362 B-scans for testing.
The underlying SS-OCT system operates at a central wavelength of 1060 nm and a 15-dB bandwidth of 75 nm.
Every B-scan contained 512 A-scans, and each original interferogram contained 1408 samples.
A second set of experiments was performed on a dataset of human retinal images, acquired with a commercial SD-OCT system (ZEISS Cirrus 5000; Carl Zeiss Meditec, Inc., Dublin, CA) with a central wavelength of 840 nm and a 10-dB bandwidth of 60 nm.
21047 B-scans from 71 volumes were used for training, 26 volumes with 8244 B-scans were reserved for testing.
Each B-scan was cropped to 224 A-scans, and each interferogram originally contained 1024 samples.
Background and DC of the spectra were removed prior to the experiments.
This dataset was used to perform additional experiments (Fig. \ref{fig:windows_plot}, bottom):
First, we selected two identical windows (gap: 0\%) from the center of the spectrum, collapsing the task to super-resolution without any complementary information while retaining the identical network architecture.
Secondly and thirdly, we adjusted the network to accept 4 and 8 windows that cover the entire spectrum in a non-overlapping (gaps: 25\%) and overlapping (gaps: 10.7\%) manner.
For these setups, only the number of input channels to the network was adjusted.
As a conventional optical baseline, we adapted frequency compounding \cite{Pircher2003Specklereductionoptical}.
For this, we matched the sizes of the two most distant (75\%) or the 8 overlapping input windows to the reference B-scans via bicubic interpolation and averaged these B-scans in linear intensity space.
The third dataset contained images of human corneas of 8 subjects, acquired using a high-resolution SD-OCT setup with a central wavelength of 800 nm and a bandwidth of 155 nm (FWHM) \cite{Schmoll2012PreciseThicknessMeasurements}.
The dataset consisted of 43 volumes with 100 B-scans each.
3600 B-scans from 6 subjects were used for training, while 700 B-scans from the remaining 2 subjects were reserved for evaluation.
Each B-scan contained 992 A-scans with 1024 samples per interferogram after resampling to remove unnecessary background areas.
All human imaging followed the tenets of the Declaration of Helsinki with informed consent obtained from all subjects.
It was approved by Salus Institutional Review Board for the retina dataset and the ethics committee of the Medical University of Vienna for the cornea dataset.



\begin{figure}[htbp]
\centering\includegraphics[width=\linewidth]{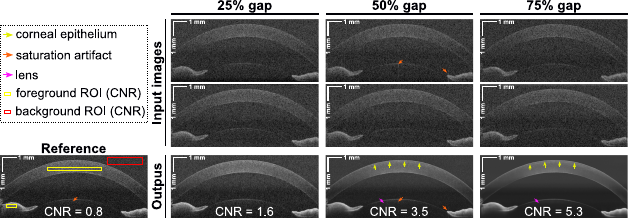}
\caption{Result from the porcine anterior segment (PASO) dataset. Images resized to match physical dimensions.}
\label{fig:results:PASO}
\end{figure}

\begin{figure*}[htbp]
\centering\includegraphics[width=\linewidth]{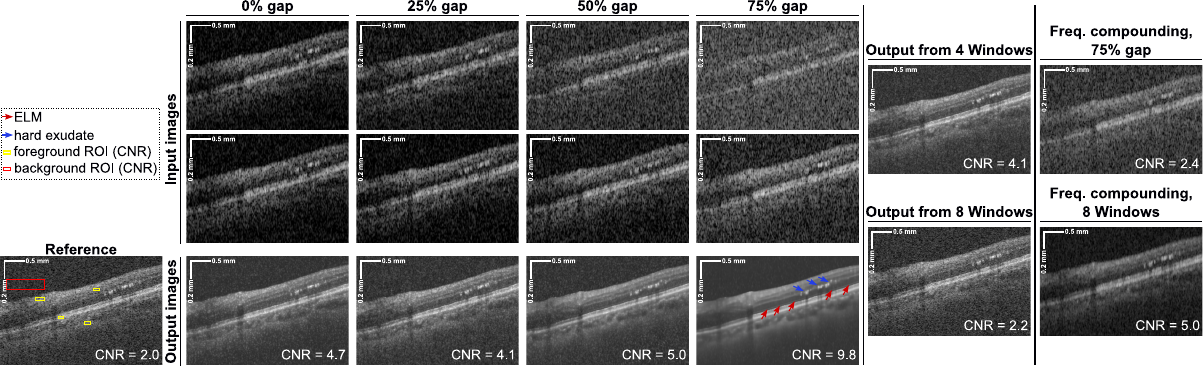}
\caption{Result from the human retina dataset, including additional experiments. Aspect ratios adjusted to enhance feature visibility.}
\label{fig:results:Cirrus}
\end{figure*}

\begin{figure}[htbp]
\centering\includegraphics[width=\linewidth]{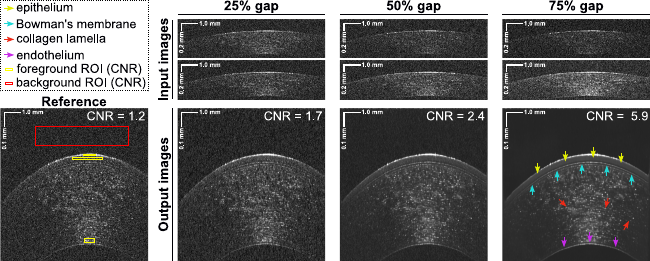}
\caption{Result from the human cornea dataset. Shown aspect ratios correspond to image sizes in pixels.}
\label{fig:results:IVC}
\end{figure}

Figures \ref{fig:results:PASO}, \ref{fig:results:Cirrus}, and \ref{fig:results:IVC} show results from all evaluated datasets at different band gaps, respectively.
For both SS- and SD-OCT data, FD-MAE successfully restored apparent resolution without introducing strong distortions at all tested band gaps.
Details that can be seen in the full-resolution reference image are successfully recovered without any anatomical priors, even when they are hardly distinguishable in the input images.
In all cases it can be observed that an increasing gap width leads to increasing noise reduction.
This was quantified by defining one background region of interest (ROI, subscript \(b\)) and \(N\) foreground ROIs (\(f_i\)) and evaluating the contrast-to-noise ratio (CNR)
\begin{equation}
	\mathrm{CNR} = \frac{1}{N} \sum_{i=1}^{N} \frac{\mu_{f_i}-\mu_b}{\sqrt{\sigma^2_{f_i}+\sigma^2_b}},
\end{equation} 
in which \(\mu\) and \(\sigma^2\) are the respective local mean and variance.
This denoising effect is present for all three datasets, albeit weaker in the retinal data (Fig. \ref{fig:results:Cirrus}), in which some background noise is still visible even for the largest band gap. 
It is visually stronger than with classical frequency compounding of 2 and even 8 windows, and accordingly has a higher CNR.

For the PASO dataset (Fig. \ref{fig:results:PASO}), recovered detail includes the thin corneal epithelium and the low-contrast lens. 
The shown example includes a weak saturation artifact in the reference image that is removed in all cases but the 50\% gap, which is the only case where it is visible in an input B-scan.
This demonstrates that for detail to be recovered, the combined input scans together must retain the underlying information.
In case of the retinal dataset (Fig. \ref{fig:results:Cirrus}), FD-MAE recovers retinal layers that not only match the reference, but even exceed it in contrast.
This includes fine, low-contrast structures such as the external limiting membrane (ELM), and pathologic structures such as hard exudates.
Frequency compounding, in contrast, suffers from visually poorer axial resolution.
In the human cornea dataset (Fig. \ref{fig:results:IVC}), structures such as the epithelium, Bowman's layer, some stromal collagen lamella, and the endothelium are clearly distinguishable, despite the lowered resolution of the input images compared to the high-resolution original.
Because this dataset has the largest overall image size, we also evaluated per-image processing, which on average finished in 34 ms, allowing video-rate processing with 29.4 B-scans per second.

Figure \ref{fig:results:Cirrus} also includes reconstruction results for the additional spectral windows tested solely on the retinal dataset.
The network can recover considerable detail even for the identical inputs (0\%), although the result in some regions appears more blurry, for example at the pathologic hard exudates.
With 4 windows that cover the entire spectrum, sharpness but also noise level increase.
At 8 overlapping windows, the network is able to visually fully recover the reference B-scans, including pixel-level noise patterns.
This underlines the suitability of the input processing and architecture to, in principle, encode and recover all information in the data.
It further demonstrates that the denoising effect observed for larger band gaps does not stem from the network or training procedure itself.


\begin{figure}[htbp]
\centering\includegraphics[width=\linewidth]{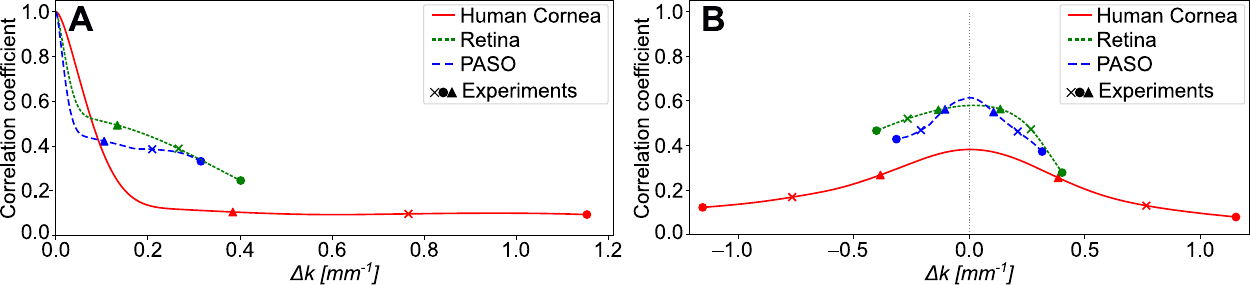}
\caption{Measured B-scan correlations for all datasets between tomograms derived from (A) two spectral windows and from (B) a spectral window and the full-bandwidth.}
\label{fig:corr_plots}
\end{figure}

To gain an understanding of the decorrelation between input images from varying gap sizes, we evaluated the Pearson cross-correlation coefficient on all three datasets (Figure \ref{fig:corr_plots} A).
It can be observed that it decreases rapidly due to decorrelation of speckle \eqref{eq:speckle_correlation} and non-speckle noise components, and also the slightly diverging axial resolutions. 
Correlation coefficients converge to non-zero plateaus, likely due to the shared structural information determined by the sample.
This convergence is in particular evident for the high-bandwidth human cornea dataset, for which input window correlation drops quickly and to a lower level compared to the PASO and retina datasets.
Via this wavenumber-dependent decorrelation, FD-MAE is related to frequency compounding \cite{Pircher2003Specklereductionoptical}, in which B-scans from different bands are averaged.

The denoising effect in FD-MAE can be explained by the noise decorrelation between input and full-bandwidth target B-scans (Fig. \ref{fig:corr_plots} B).
For this analysis, image sizes were matched via bilinear upsampling of the low-resolution input B-scans.
According to the Noise2Noise-principle \cite{Lehtinen2018Noise2NoiseARTICLE}, independence of noise patterns between training inputs and targets yields noise-reduced predictions.
In FD-MAE, such noise decoupling is approximately achieved when the introduced band gap is sufficiently large.
It thereby serves as a self-supervised denoising approach, in the sense that it only requires single OCT interferograms with sufficient bandwidth for training.
The measured decorrelation is less pronounced for the Cirrus dataset especially towards negative \(\Delta k\), which conversely exhibits a comparably slightly weaker denoising effect (cf. Fig. \ref{fig:results:Cirrus}).

\begin{figure}[htbp]
\centering\includegraphics[width=\linewidth]{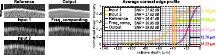} 
\caption{Comparison of normalized rising edges at a corneal surface in inputs, reference, and output.}
\label{fig:CorneaAxialRes}
\end{figure}

To evaluate detail and resolution recovery, we investigated the surface profiles along a flat corneal surface stretch from the PASO dataset (Fig. \ref{fig:CorneaAxialRes}, left) for the largest possible band gap.
While it does not allow direct assessment of the axial resolution, which would be expected to be degraded by a factor of 4, the corneal surface represents a rapid transition which here serves as a substitute for e.g. a mirror measurement.
Due to its asymmetry, we only consider the rising edge and analyze the rise from half to full peak intensity (half-width of half maximum, HWHM).
For evaluation, we aligned the 140 individual A-scans by peak location to obtain average profiles of the rising corneal edge (Fig. \ref{fig:CorneaAxialRes}, right).
Low-bandwidth inputs from the ends of the spectrum yielded HWHMs of 37.82 µm and 40.06 µm.
Frequency compounding approximately preserved this resolution, reaching an HWHM of 40.11 µm.
After reconstruction by the FD-MAE, the average HWHM was 16.23 µm, which closely matches the reference width of 15.7 µm.

We further calculated the signal-to-noise ratio
\begin{equation}
	\mathrm{SNR} = 10\log_{10} \left( \frac{ \mu_{\mathrm{peak}} - \mu_{\mathrm{noise}}}{\sigma^2_{\mathrm{noise}}} \right) [dB]
\end{equation}
in which \(\mu_{\mathrm{peak}}\) is the average linear peak intensity while \(\mu_{\mathrm{noise}}\) and \(\sigma^2_{\mathrm{noise}}\) are mean and variance of the background noise above the corneal surface.
SNR in low-resolution input scans (27.92 dB and 27.52 dB) was slightly higher than in the reference scan (26.41 dB) due to increased blurring of the background noise.
Frequency compounding from the two input windows improved the SNR to 30.95 dB.
After reconstruction by the FD-MAE, the SNR was greatly improved to 59.82 dB, as background noise was almost fully suppressed.
Furthermore, it can again be qualitatively observed that FD-MAE also recovered the corneal epithelium right below the corneal surface.
Its thickness visually corresponds well to that seen in the reference scan.

In summary, our proposed FD-MAE provides a flexible foundation for combining multiple lasers in FD-OCT, being applicable at different spectral gaps and variable numbers of sources.
Furthermore, with the inherent denoising characteristic that increases with the band gap, this method can not only achieve, but even surpass the image quality obtained from single-source OCT systems.
Compared to classical frequency compounding \cite{Pircher2003Specklereductionoptical}, FD-MAE not only leads to higher SNR improvements, but it is also edge-preserving and can yield the axial resolution of a combined (broadband) acquisition.
To optimize the achievable image quality, the SNR within the individual spectral ranges should be optimized, as the network can only recover information that is contained in the combined set of input images.
Even for broadband systems, training with artificial gaps can serve as an accessible self-supervised denoising approach, only requiring interferograms from single acquisitions for training.
The proposed method may further offer opportunities for compressed sensing, reducing data storage demands by acquiring only part of the spectra and reconstructing high-quality images using a pretrained network.

\footnotesize

\paragraph*{Funding:} This work was funded by ZEISS.

\paragraph*{Acknowledgments:} This article expands on a prior conference contribution \cite{Nienhaus2026NetworkbasedOCTARTICLE}. 
We thank Conor Leahy for feedback and discussion, and ZEISS for providing access to the retinal dataset. 

\paragraph*{Data Availability:} The PASO dataset is publicly available \cite{Nienhaus2025PASOMultipurposePorcineARTICLE}. All other data are not publicly available due to privacy concerns.


\printbibliography

\end{document}